\documentclass[10pt,letterpaper]{article}
\usepackage[top=0.85in,left=2.75in,footskip=0.75in]{geometry}

\usepackage{changepage}

\usepackage[utf8]{inputenc}

\usepackage{textcomp,marvosym}

\usepackage{fixltx2e}

\usepackage{amsmath,amssymb}

\usepackage{cite}
\usepackage{float}
\usepackage{nameref,hyperref}

\usepackage[right]{lineno}

\usepackage{microtype}
\DisableLigatures[f]{encoding = *, family = * }

\usepackage{rotating}
\usepackage{slashbox}
\raggedright
\usepackage[aboveskip=1pt,labelfont=bf,labelsep=period,justification=raggedright,singlelinecheck=off]{caption}

\makeatletter
\renewcommand{\@biblabel}[1]{\quad#1.}
\makeatother

\date{}

\usepackage{lastpage,fancyhdr,graphicx}
\usepackage{epstopdf}
\fancyheadoffset[L]{2.25in}
\fancyfootoffset[L]{2.25in}
\newcommand{\acc}{$\mathit{Accuracy}$}
\newcommand{\accone}{$\mathit{Accuracy}\!\pm\!1$}

\begin{document}
\vspace*{0.35in}

% Title must be 250 characters or less.
% Please capitalize all terms in the title except conjunctions, prepositions, and articles.
\begin{flushleft}
{\Large
\textbf\newline{Emotional Dynamics in the Age of Misinformation}
}
\newline
% Insert author names, affiliations and corresponding author email (do not include titles, positions, or degrees).
\\
Fabiana Zollo\textsuperscript{1},
Petra Kralj Novak\textsuperscript{2},
Michela Del Vicario\textsuperscript{1},
Alessandro Bessi\textsuperscript{1,3},
Igor Mozeti\v{c}\textsuperscript{2},
Antonio Scala\textsuperscript{4},
Guido Caldarelli\textsuperscript{1},
Walter Quattrociocchi\textsuperscript{1,*}
\\
\bigskip
\bf{1} IMT Institute for Advanced Studies, Lucca, Italy
\\
\bf{2} Jo\v{z}ef Stefan Institute, Ljubljana, Slovenia
\\
\bf{3} IUSS, Pavia, Italy
\\
\bf{4} ISC-CNR, Rome, Italy
\bigskip

% Insert additional author notes using the symbols described below. Insert symbol callouts after author names as necessary.
% 
% Remove or comment out the author notes below if they aren't used.
%
% Primary Equal Contribution Note
%\Yinyang These authors contributed equally to this work.

% Additional Equal Contribution Note
% Also use this double-dagger symbol for special authorship notes, such as senior authorship.
%\ddag These authors also contributed equally to this work.
 
% Current address notes
%\textcurrency a Insert current address of first author with an address update
% \textcurrency b Insert current address of second author with an address update
% \textcurrency c Insert current address of third author with an address update

% Deceased author note
%\dag Deceased

% Group/Consortium Author Note
%\textpilcrow Membership list can be found in the Acknowledgments section.

% Use the asterisk to denote corresponding authorship and provide email address in note below.
* walterquattrociocchi@gmail.com

\end{flushleft}
% Please keep the abstract below 300 words
\section*{Abstract}
According to the World Economic Forum, the diffusion of unsubstantiated rumors on online social media is one of the main threats for our society.

The disintermediated paradigm of content production and consumption on online social media might foster the formation of homophile communities (echo-chambers) around specific worldviews.
Such a scenario has been shown to be a vivid environment for the diffusion of false claims, in particular with respect to conspiracy theories. Not rarely, viral phenomena trigger naive (and funny) social responses -- e.g., the recent case of Jade Helm 15 where a simple military exercise turned out to be perceived as the beginning of the civil war in the US.
In this work, we address the emotional dynamics of collective debates around distinct kind of news -- i.e., science and conspiracy news -- and inside and across their respective polarized communities (science and conspiracy news).

Our findings show that comments on conspiracy posts tend to be more negative than on science posts. However, the more the engagement of users, the more they tend to negative commenting (both on science and conspiracy). Finally, zooming in at the interaction among polarized communities, we find a general negative pattern.  As the number of comments increases -- i.e., the discussion becomes longer -- the sentiment of the post is more and more negative.
% Please keep the Author Summary between 150 and 200 words
% Use first person. PLOS ONE authors please skip this step. 
% Author Summary not valid for PLOS ONE submissions.   
%\section*{Author Summary}

%\linenumbers

\section*{Introduction}

People online get informed, discuss and shape their opinions~\cite{loader2014networked,scott2014entanglements,magro2012review}. 
Indeed, microblogging platforms such as Facebook and Twitter allow for the direct and disintermediated production and consumption of contents~\cite{farley2014happens,brabazon2015digital,meraz2009there,friggeri2014rumor}. 
The information heterogeneity might facilitate users selective exposure to specific content and hence the aggregation in homophile communities ~\cite{colleoni2014echo,yardi2010dynamic,quattrociocchi2014opinion,adamic2005political,quattrociocchi2010simulating,josang2011taste,quattrociocchi2011opinions,anagnostopoulos2014viral}. 
In such echo-chambers users interaction with different narratives is reduced and the resulting debates are often polarized (misinformation)~\cite{Mocanu2015,bessi2014economy,bessi2015trend,bessi2014science,bessi2014social}.

Unfortunately, despite the enthusiastic rhetoric about {\em collective intelligence}~\cite{levy1997collective,bonabeau2009decisions,surowiecki2005wisdom}, the direct and undifferentiated access to the knowledge production process is causing opposite effects -- e.g., the recent case of Jade Helm 15 \cite{JadeHelm} where a simple military exercise turned out to be perceived as the beginning of the civil war in the US.
Unsubstantiated rumors often jump the credulity barrier and trigger naive social responses.
To an extent that, recently, the World Economic Forum labeled {\em massive digital misinformation} as one of the main threats to our society. 
Individuals may be uninformed or misinformed, and the debunking campaigns against unsubstantiated rumors do not seem to be effective~\cite{Kuklinski2000}.

Indeed, the factors behind the acceptance of a claim (whether substantiated or not) may be altered by normative social influence or by the coherence with the system of beliefs of the individual~\cite{Zhu2010,Loftus2011,byford2011conspiracy,finerumor,hogg2011extremism}, making the preferential driver of contents the {\em confirmation bias} -- i.e., the tendency to select and interpret information coherently with one's system of beliefs.

In~\cite{bessi2014science,Mocanu2015,bessi2014economy} it has been pointed out that the more users are exposed to unsubstantiated rumors, the more they are likely to jump the credulity barrier.  

In this work we analyze a collection of {\em conspiracy} and {\em scientific} news sources in the Italian Facebook over a time span of four years. 
The main distinctive feature of the two categories of pages is the possibility to verify the reported content. Scientific news are generally fact-checked and are the results of a peer review process. Conversely, conspiracy news are generally partial information about a secret plot.
We identify pages diffusing conspiracy news -- i.e., pages promoting contents {\em neglected} by main stream media and scientific pages -- aiming at diffusing scientific results.
To have an exhaustive list of pages, we define the space of our investigation with the help of Facebook groups very active in debunking conspiracy stories and unverified rumors ({\em Protesi di Complotto}, {\em Che vuol dire reale}, {\em La menzogna diventa verit\`a e passa alla storia}).

We target emotional dynamics inside and across content polarized communities. 
In particular, we apply sentiment analysis techniques to the comments
of the Facebook posts, and study the aggregated sentiment with respect to scientific and conspiracy-like information.
The sentiment analysis is based on a supervised machine learning approach,
where we first annotate a large sample of comments, and then build
a Support Vector Machine (SVM~\cite{vapnik95}) classification model. 
The model is then applied to associate each comment with one sentiment value:
negative, neutral, or positive. The sentiment is intended to express
the emotional attitude of Facebook users when posting comments.

Although other studies apply sentiment analysis to social media~\cite{pak2010twitter,go2009twitter,tumasjan2010predicting,luong2015public}, our work is the first linking the interplay between communities emerging around shared narratives and specifically addressing the emotional dynamics with respect to misinformation spreading.
Indeed, this work provides important insights toward the understanding of the social factors behind contents consumption and the formation of polarized and homophile clusters with a specific interest in conspiracy-like information.

We focus on the emotional behavior of about 280k Facebook Italian users and through a thorough quantitative analysis, we find that the sentiment on conspiracy pages tends to be more negative than that on science pages. 
In addition, by focusing on polarized users -- i.e., users mainly exposed to one specific content type (science or conspiracy) -- we capture an overall increase of the negativity of the sentiment. 
According to our results, the more active polarized users are, the more they tend to be negative, both on science and conspiracy. 
Furthermore, the sentiment of polarized users is negative also when they interact with one another. 
Also, as the number of comments increases -- i.e., the discussion turns longer-- the sentiment is more and more negative.

\section*{Results and Discussion}

\subsection*{Sentiment classification}

Emotional attitude towards different topics can be roughly approximated by the
sentiment expressed in texts. It is difficult to exactly formalize the
sentiment metrics since there are often disagreements between humans,
and even individuals are not consistent with themselves. 

In this study, as is often in the sentiment analysis literature~\cite{pang08}, 
we have approximated the sentiment with an ordinal scale of three values:
\textit{negative} ($-$), \textit{neutral} ($0$), and \textit{positive} ($+$).
Even with this rough approximation, and disagreements on single cases, it turns out that on a large scale, when one deals with thousands of sentiment assignments, the aggregated sentiment converges to stable values \cite{pang2002thumbs}.

Our approach to automatic sentiment classification of texts is based
on supervised machine learning. There are four steps:
(i) a sample of texts is manually annotated with sentiment,
(ii) the labeled set is used to train and tune a classifier,
(iii) the classifier is evaluated on an independent test set, and
(iv) the classifier is applied to the whole set of texts.

We have collected over one million of Facebook comments
(see Table~\ref{tab2} for details). 
About 20k were randomly selected for manual annotation.
We have engaged 22 native Italian speakers, active on Facebook,
to manually annotate the comments by sentiment.
The annotation is supported by a web-based platform 
Goldfinch\footnote{provided by Sowa Labs http://www.sowalabs.com},
and was accomplished in two months.
About 20\% of comments were intentionally duplicated, in order to
measure the mutual (dis)agreement of human annotators.

There are several metrics to evaluate the inter-annotator agreement
and performance of classification models.
We argue that the inter-annotator agreement provides an upper bound
that the best classification model can achieve. In practice, however,
different learning algorithms have various limitations, and, most
importantly, only a limited amount of training data is available.
In order to compare the classifier performance to the inter-annotator
agreement, we have selected three metrics used for both:
\acc, $\overline{F_{1}}$, and \accone\, (see details in 
the Methods section). 
\acc\, is the fraction of correctly classified
examples for all three sentiment classes. 
$F_{1}$ is the harmonic mean of precision and recall for each class.
$\overline{F_{1}}(-,+)$ is the average of $F_{1}$ for the negative and 
positive class only.
It is a standard measure of performance for sentiment classification~\cite{kiritchenko2014sentiment}.
The idea is that the misclassification of neutral sentiment can be ignored
as it is less important then the extremes, i.e., negative or positive sentiment
(however, it still affects their precision and recall).
\accone\, completely ignores the neutral class.
It counts as errors just the negative sentiment examples predicted as
positive, and vice versa.
Table~\ref{tab0} gives the results. 
One can see that the classifier has reached
performance close to human agreement, and that there is still some
room for improvement in terms of \acc\, and $\overline{F_{1}}$. However,
in terms of extreme errors, i.e., $1 - $\accone\, the performance of 
the classifier is as good as the agreement between the annotators.

\begin{table}[h]
\caption{ \textbf{Comparison of the inter-annotator agreement and classifier
        performance over three evaluation metrics.}}
	\begin{tabular}{|l|c|c|}
		\hline\bf {  }  & \bf {Annotator agreement} & \bf {Sentiment classifier}  \\ 
		\hline
		No. of testing examples & $ 3,262 $  & $ 3,928 $\\ \hline
		\acc($-,0,+$)       & $ 72\% $ & $ 64\% $\\ \hline
		$\overline{F_{1}}(-,+)$ & $ 73\% $ & $ 65\% $ \\ \hline
		\accone($-,+$)     & $ 97\% $ & $ 97\% $ \\ \hline
	\end{tabular}\newline
\label{tab0}
\end{table}

\begin{figure}[H]
\includegraphics[width=0.55\textwidth]{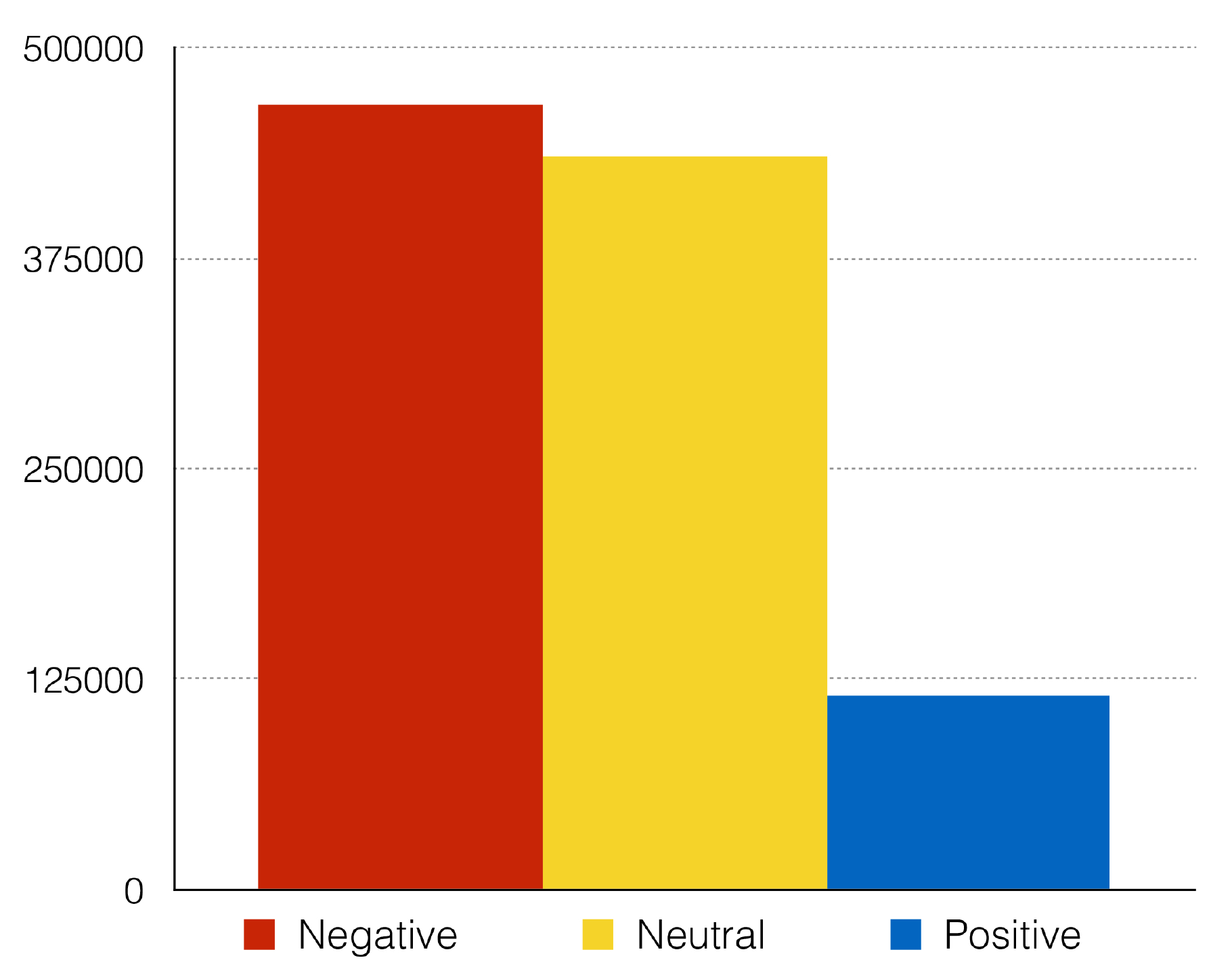}
\caption{{\bf Sentiment distribution over the entire set of 1 million comments.}}
\label{fig0}
\end{figure}

Fig.~\ref{fig0} gives the distribution of sentiment values after
applying the classification model to the entire set of over 1 million
comments.
We assume that the sentiment values are ordered, and that the
difference from the neutral value to both extremes, 
negative and positive, is the same.
Thus one can map the sentiment values from ordinal to a real-valued interval
$[-1, +1]$. The average sentiment over the entire set
is $-0.34$, prevailingly negative.

\subsection*{Sentiment on science and conspiracy posts}

\begin{figure}[H]
\includegraphics[width=\textwidth]{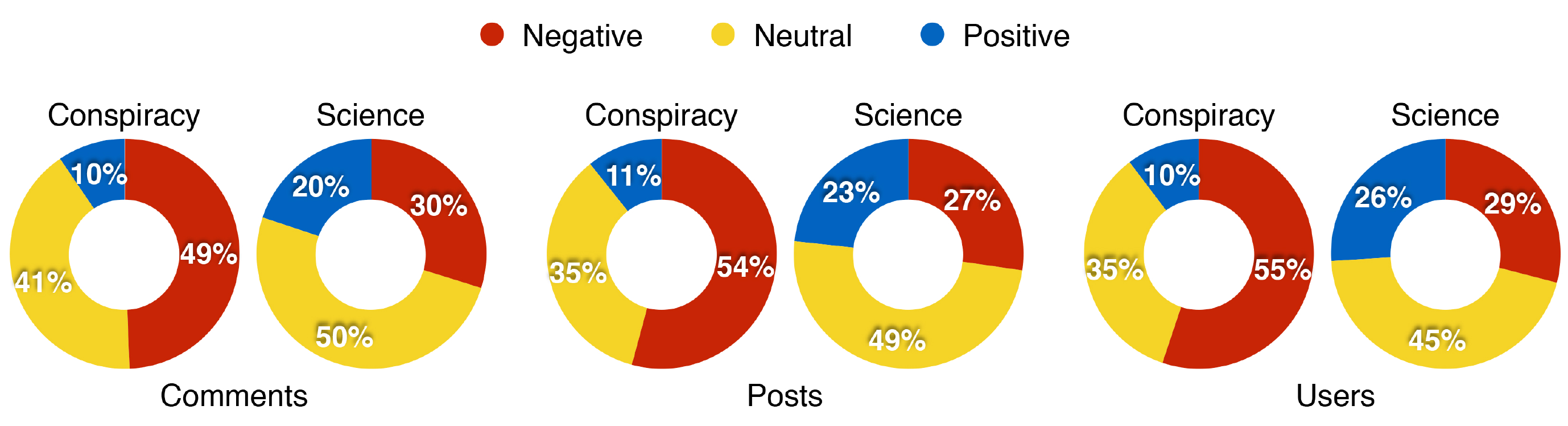}
\caption{{\bf Sentiment on science and conspiracy pages.} Proportions of negative, neutral and positive comments (\textit{left}), posts (\textit{center}), and users (\textit{right}) both on science and conspiracy pages.}
\label{fig1}
\end{figure}

The sentiment analysis and classification task allowed us to associate each comment of our dataset to a sentiment value -- i.e., $-1$ if \textit{negative}, $0$ if \textit{neutral}, and $1$ if \textit{positive}. 
Taking all the comments of science and conspiracy posts, we can simply divide them into negative, neutral and positive (Fig.~\ref{fig1}, \textit{left}), and analyze their proportions. 
We find that the majority of comments on science pages ($70\%$) is neutral or positive, differently from conspiracy pages ($51\%$). Moreover, comments on science pages are twice as positive ($20\%$) than those on conspiracy pages ($10\%$). 

To measure the effect induced on users by a post, we compute the average sentiment of all its comments.
We grouped posts sentiment by defining three thresholds; in particular, we say a post to be \textit{negative} if the average sentiment $\in [-1, -0.3]$, \textit{neutral} if $\in (-0.3, 0.3)$, and \textit{positive} if  $\in [0.3, 1]$. 
Fig.~\ref{fig1} \textit{(center)} shows the aggregated sentiment of science and conspiracy posts. Notice that the sentiment of conspiracy posts is mainly negative ($54\%$), differently from science posts, for which the negative sentiment represents only the $27\%$. On the other hand, it is twice as positive for science posts ($23\%$) than for conspiracy posts ($11\%$). 

When focusing on users, the approach is analogous. 
We define the sentiment of a user as the mean of the sentiment of all her comments. The mean sentiment for each user is then classified as negative, neutral, or positive by means of the same thresholds used for posts.  
Fig.~\ref{fig1} \textit{(right)} shows the aggregated sentiment both for science and conspiracy users. We find that the sentiment of users commenting on conspiracy pages is mainly negative ($55\%$), while the sentiment of a small fraction of users ($10\%$) is positive. 
On the contrary, the sentiment of users commenting on science pages is particularly neutral ($45\%$), and negative only for $29\%$ of users. Almost the same percentage ($26\%$) is represented by positive sentiment.

\subsection*{Sentiment and virality}

\begin{figure}[H]
\includegraphics[width=0.9\textwidth]{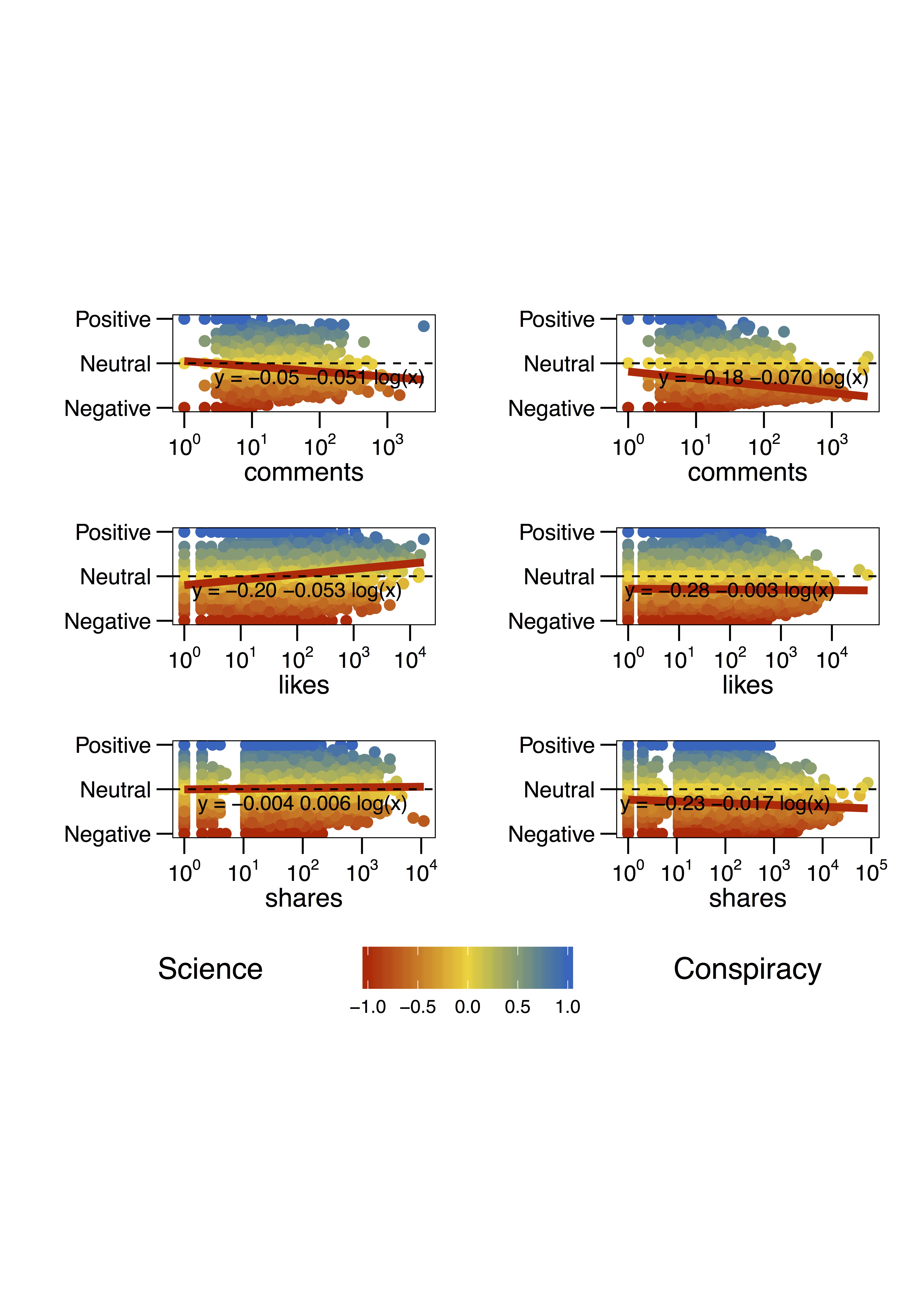}
\caption{{\bf Sentiment and post consumption.} Aggregated sentiment of posts as a function of their number of comments, likes, and shares, both for science (\textit{left}) and conspiracy (\textit{right}). Negative (respectively, neutral, positive) sentiment is denoted by red (respectively, yellow, blue) color. The sentiment has been regressed w.r.t. the logarithm of the number of comments/likes/shares.}
\label{fig2}
\end{figure}

Now we focus on the interplay between the virality of a post and its generated sentiment. 
In particular we want to understand how the sentiment varies for increasing levels of comments, likes, and shares. Notice that each of these actions has a particular meaning~\cite{Ellison2007,Joinson:2008,Viswanath:2009}. A {\em like} stands for a positive feedback to the post; a {\em share} expresses the will to increase the visibility of a given information; and a {\em comment} is the way in which online collective debates take form around the topic promoted by posts. Comments may contain negative or positive feedbacks with respect to the post. Fig.~\ref{fig2} shows the aggregated sentiment of a post as a function of its number of comments \textit{(top)}, likes \textit{(center)}, and shares \textit{(bottom)} both for science \textit{(left)} and conspiracy \textit{(right)} posts. 
The sentiment has been regressed w.r.t. the logarithm of the number of comments (resp., likes, shares)\footnote{
We do not show confidence intervals, since they are defined (C.I. 95\%) as 
$\bar{X} \pm S.E. = \bar{X} \pm 1.96 \frac{\sigma}{\sqrt{n}}$ and when $n \to \infty$, $S.E. = 0$.
}.
We notice that the sentiment decreases both for science and conspiracy when the number of comments of the post increases. 
However, we also note that it becomes more positive for science posts when the number of likes and shares increase, differently from conspiracy posts.

To assess the direct relationship between the number of comments and the negativity of the sentiment, a randomization test was performed. In particular, we took all the comments of science (resp., conspiracy) posts and randomly reassigned the original sentiments. Then, we regressed the sentiment w.r.t. the number of comments and compared the obtained slope with the one shown in Fig.~\ref{fig2}\textit{(top)}. Over 10k randomized tests, the obtained slope was always greater than the original one. Therefore, given that the negative relationship between the sentiment and the length of the discussion disappears when the comment sentiments are randomized, we conclude that the length of the discussion is a relevant dimension when considering the negativity of the sentiment.

%%%%%% JPEG FOR CONVENIENCE, USE PDF FOR THE VERY FINAL VERSION %%%%%

Summarizing, we found that both comments and posts, as well as users of conspiracy pages tend to be much more negative than those of science pages. Interestingly, the sentiment becomes more and more negative when the number of comments of the post increases -- i.e., the discussion becomes longer-- both on science and conspiracy pages. 
However, differently from conspiracy posts, when the number of likes and shares increases, the aggregated sentiment of science posts becomes more and more positive.

\subsection*{Sentiment and users activity}

In this section we aim at understanding more in depth how the sentiment changes with respect to users' engagement in one of the two communities. 
Previous works~\cite{bessi2014science, bessi2014economy, bessi2014social} showed that the distribution of the users activity on the different contents is highly polarized. Therefore we now want to focus on the sentiment of polarized users. 
More precisely, we say a user to be polarized on science (respectively, on conspiracy) if she left more than $95\%$ of her likes on science (respectively, on conspiracy) posts (for further details about the effect of the thresholding refer to the Methods Section). 

\begin{figure}[H]
\includegraphics[width=0.8\textwidth]{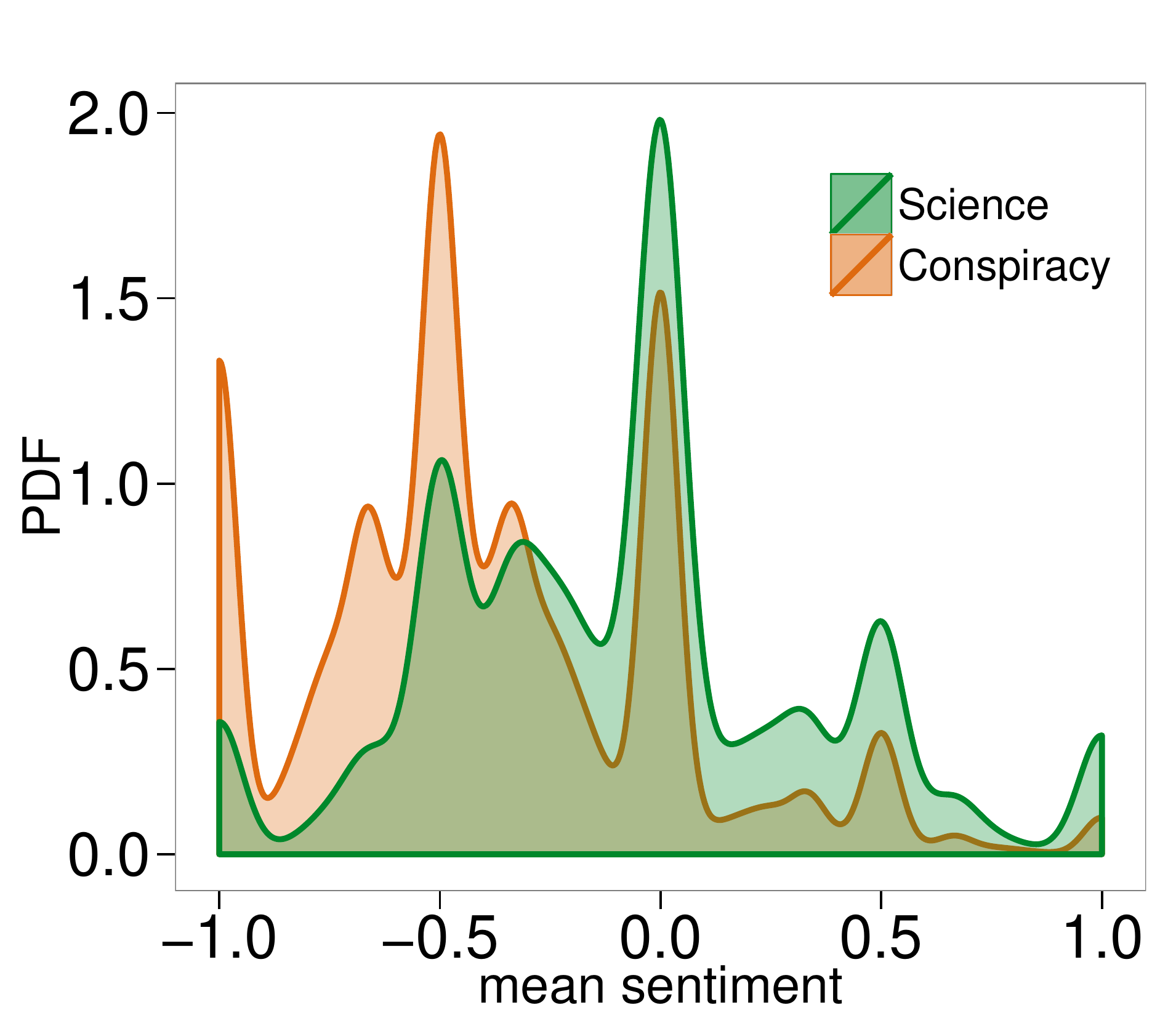}
\caption{{\bf Sentiment and polarization.} Probability Density Function (PDF) of the mean sentiment of polarized users having commented at least twice, where $-1$ corresponds to negative sentiment, $0$ to neutral and $1$ to positive.}
\label{fig3}
\end{figure}

Therefore, we take all polarized users having commented at least twice, i.e., $14,887$ out of $33,268$ users polarized on science and $67,271$ out of $135,427$ users polarized on conspiracy. Fig.~\ref{fig3} shows the Probability Density Function (PDF) of the mean sentiment of polarized users with at least two comments. 
In Table~\ref{tab1} we compare the mean sentiment of all users and polarized users having commented at least twice. 
Our results show that the overall negativity increases w.r.t. all users, such a feature is more evident on the conspiracy side.

\begin{table}[H]
\caption{
{\bf Sentiment and polarized users.}}
\begin{tabular}{|l|c|c|c|c|c|cl|l|}
\hline
{}& \multicolumn{2}{|c|}{\bf Science} & \multicolumn{2}{|c|}{\bf Conspiracy}\\ \hline
 {\bf Sentiment} & {\bf All users} & {\bf Polarized} & {\bf All users} & {\bf Polarized} \\ \hline
\textit{Negative} & 29\% & 34\%  & 55\% & 66\% \\ \hline
\textit{Neutral} & 45\% & 46\% &  35\% & 27\% \\ \hline
\textit{Positive} & 26\% & 20\% & 10\% & 7\% \\ \hline
\end{tabular}
\begin{flushleft} Mean sentiment of all users and polarized users having commented at least twice.\end{flushleft}
\label{tab1}
\end{table}

\begin{figure}[H]
\includegraphics[width=0.8\textwidth]{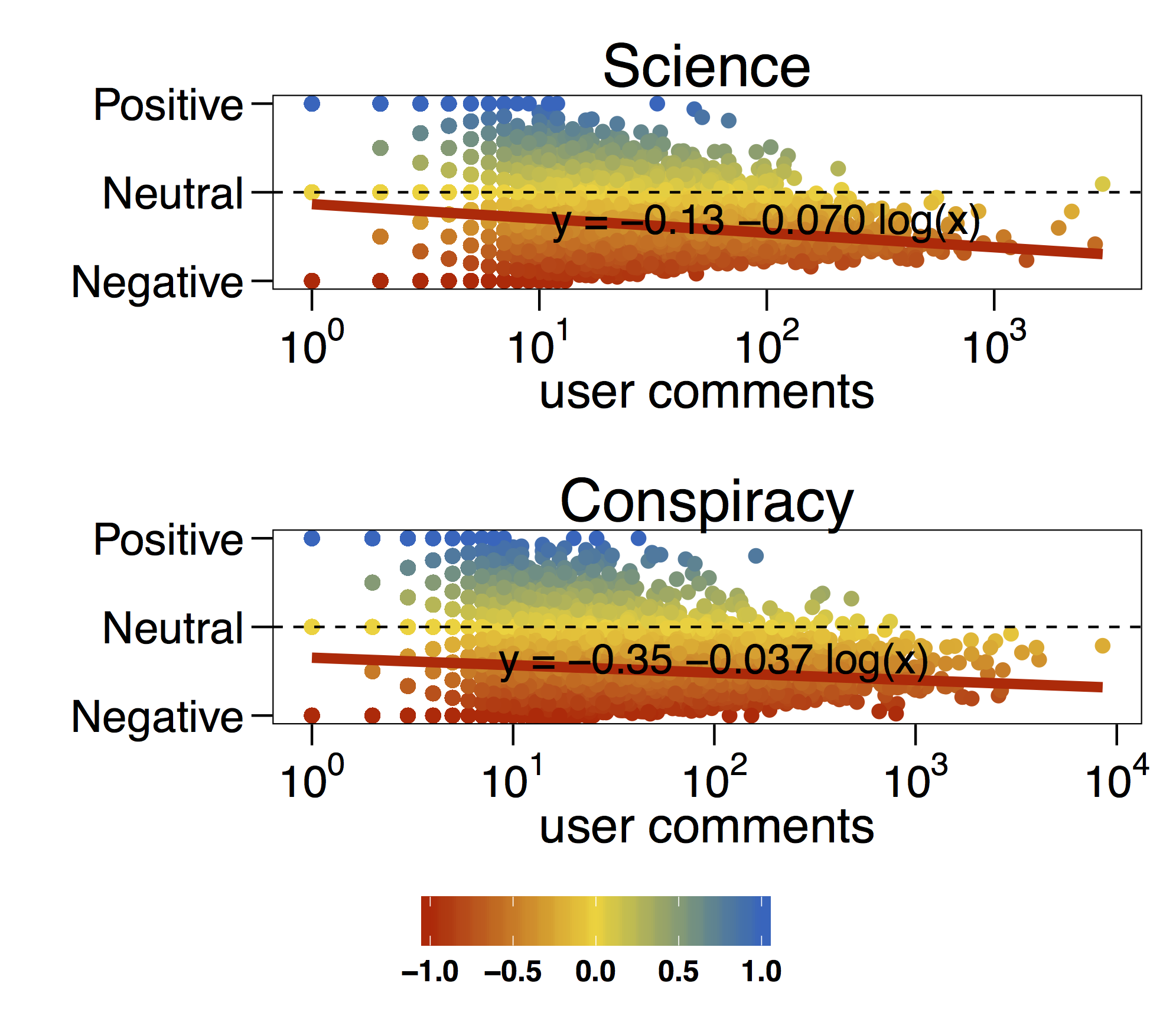}
\caption{{\bf Sentiment and commenting activity.} Average sentiment of polarized users as a function of their number of comments. Negative (respectively, neutral, positive) sentiment is denoted by red (respectively, yellow, blue) color. The sentiment has been regressed w.r.t. the logarithm of the number of comments.}
\label{fig4}
\end{figure}

We now want to investigate how the mean sentiment of a user changes with respect to her commenting activity --i.e., when her total number of comments increases. 
In Fig.~\ref{fig4} we show the mean sentiment of polarized users as a function of their number of comments. The more active a polarized user is, the more she tends toward negative values both on science and conspiracy posts. 
The sentiment has been regressed w.r.t. the logarithm of the number of comments. Interestingly, the sentiment of science users decreases faster than that of conspiracy users. 
We performed a randomization test taking all comments on both categories and then randomly reassigning the original sentiments. 
Then, we regressed the sentiment w.r.t. the number of comments and compared the obtained slope with the one shown in Fig.~\ref{fig4}. 
The obtained slope over 10k randomized tests was always greater than the original one. Therefore users activity is a relevant dimension when considering the value of the sentiment, which is more and more negative on both categories when the users activity increases.

% Please do not create a heading level below \subsection. For 3rd level headings, use \paragraph{}. 
\subsection*{Interaction across communities}

In this section we aim at investigating the sentiment when usual consumers of science and conspiracy news meet. 
To do this we pick all  posts representing the arena where the debate between science and conspiracy users takes place. In particular, we select all posts commented at least once by both a user polarized on science and a user polarized on conspiracy. 
We find $7,751$ such posts (out of $315,567$) --reinforcing the fact that the two communities of users are strictly separated and do not often interact with one another. 

\begin{figure}[H]
\begin{center}
\includegraphics[width=0.5\textwidth]{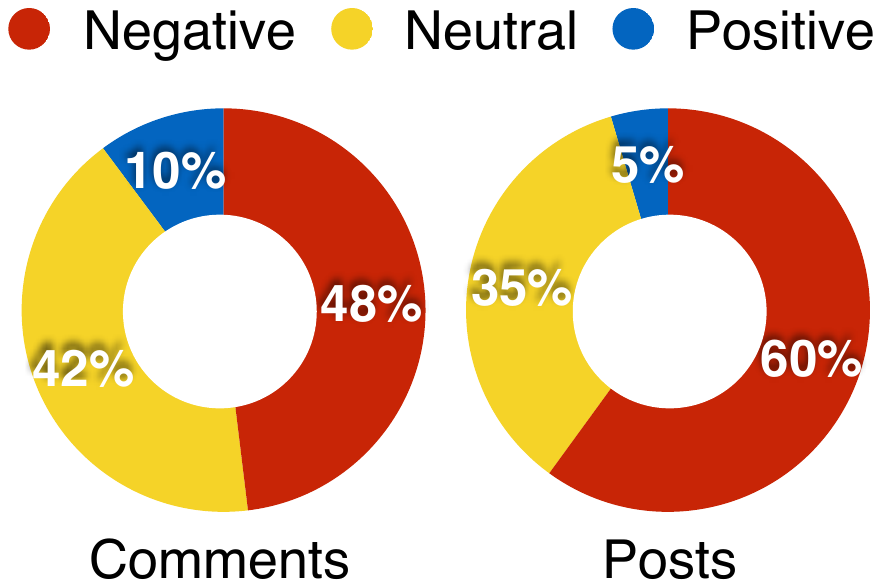}
\caption{{\bf Sentiment between communities.}
Proportions of negative, neutral, and positive comments (\textit{left}) and posts (\textit{right}) of all the posts commented at least once by both a user polarized on science and a user polarized on conspiracy.}
\label{fig5}
\end{center}
\end{figure}

\begin{figure}[H]
\includegraphics[width=0.8\textwidth]{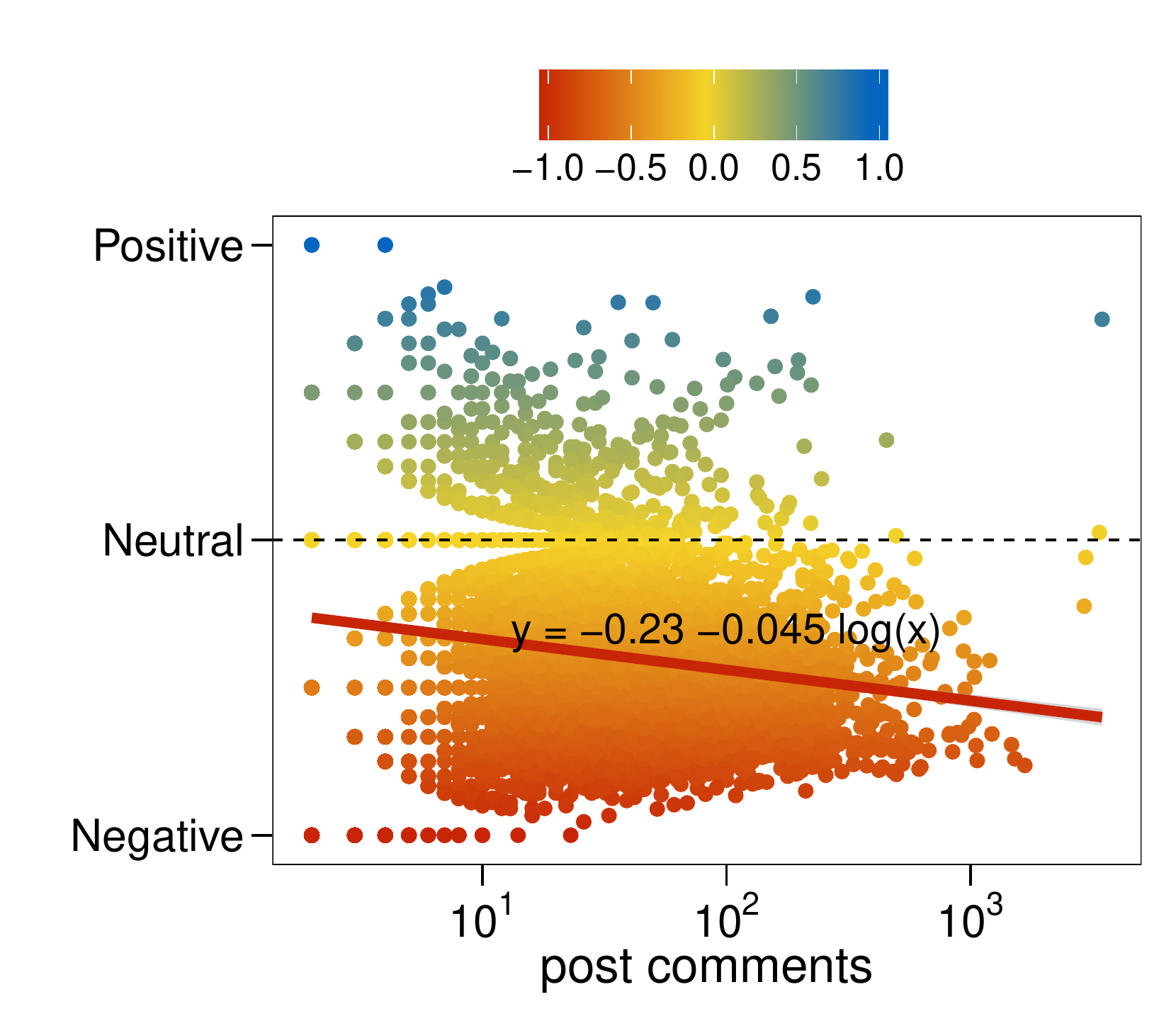}
\caption{{\bf Sentiment and discussion.}
Aggregated sentiment of posts as a function of their number of comments. Negative (respectively, neutral, positive) sentiment is denoted by red (respectively,  yellow, blue) color.}
\label{fig6}
\end{figure}

In Fig.~\ref{fig5} we show the proportions of negative, neutral, and positive comments  (\textit{left}) and posts (\textit{right}). 
The aggregated sentiment of such posts is slightly more negative ($60\%$) than for general posts ($54\%$ for conspiracy, $27\%$ for science, see Fig.~\ref{fig1}). 
When focusing on comments, we have similar percentages of neutral ($42\%$) and negative ($48\%$) comments, while a small part ($10\%$) is represented by positive comments. 
We want to understand if the sentiment correlates with the length of the discussion. 
Hence, we analyze how the sentiment changes when the number of comments of the post increases, as we previously did for \textit{general} posts (Fig.~\ref{fig2}). 
Fig.~\ref{fig6} shows the aggregated sentiment of such posts as a function of their number of comments. 
Clearly, as the number of comments increases --i.e., the discussion becomes longer-- the sentiment is more and more negative. 
Moreover, comparing with Fig.~\ref{fig2}, when communities interact with one another, posts show a higher concentration of negative sentiment.

Also in this case we performed a randomization test taking all the comments and randomly reassigning the original sentiments. Then, we regressed the sentiment w.r.t. the number of comments and compared the obtained slope with the one shown in Fig.~\ref{fig5}. Over 10k randomized tests, the obtained slope was always greater than the original one. Therefore, we conclude that the length of the discussion does affect the negativity of the sentiment.

\section*{Conclusions}

In this work we analyzed the emotional dynamics on pages of opposite worldviews, science and conspiracy. Previous works~\cite{bessi2014science, bessi2014economy, bessi2014social} showed that users are strongly polarized towards the two narratives. Moreover, we found that users of both categories seem to not distinguish between verified contents and unintentional false claims. In this manuscript we focused on the emotional behavior of the same users on Facebook. In general, we noticed that the sentiment on conspiracy pages tends to be more negative than that on science pages. In addition, by focusing on polarized users, we identified an overall increase of the negativity of the sentiment. In particular, the more active polarized users, the more they tend to be negative, both on science and conspiracy. Furthermore, the sentiment of polarized users is negative also when they interact with one another. Also in this case,  as the number of comments increases --i.e., the discussion becomes longer-- the sentiment of the post is more and more negative.

\section*{Acknowledgments}
\nolinenumbers

Funding for this work was provided by the EU FET project MULTIPLEX, no. $317532$, SIMPOL, no. $610704$, the FET project DOLFINS 640772 (H2020), SoBigData 654024 (H2020), and CoeGSS 676547 (H2020). The funders had no role in study design, data collection and analysis, decision to publish, or preparation of the manuscript. 
We want to thank Alessandro Damiano Sabatino, Alessandro Mattedi, Andrea Nardinocchi, Angelo Bonaldo, Duccio Gamannossi degl'Innocenti, Fabio Mignogna, Giacomo Sorbi, Giulia Borrione, Ilaria Zanardi, Ilaria Zanetti, Lorenzo Stella, Michele Degani, Nicola Filardi, Nicola Palo, Paolo Camboni, Salvatore Previti, and Stefano Alpi for their precious suggestions and help in annotating the dataset for the sentiment classification task.

\section*{Methods}

\subsection*{Ethics statement}
The entire data collection process has been carried out exclusively through the Facebook Graph API~\cite{fb_graph_api}, which is publicly available, and for the analysis (according to the specification settings of the API) we used only public available data (users with privacy restrictions are not included in the dataset). The pages from which we download data are public Facebook entities (can be accessed by anyone). User content contributing to such pages is also public unless the user's privacy settings specify otherwise and in that case it is not available to us.

\subsection*{Data collection}

We identified two main categories of pages: conspiracy news --i.e. pages promoting contents neglected by main stream media-- and science news. The first category includes all pages diffusing conspiracy information --pages which disseminate controversial information, most often lacking supporting evidence and sometimes contradictory to the official news (i.e., conspiracy theories). The second category is that of scientific dissemination, including scientific institutions and scientific press having the main mission to diffuse scientific knowledge. We defined the space of our investigation with the help of Facebook groups very active in debunking conspiracy theses (\textit{Protesi di Complotto}, \textit{Che vuol dire reale}, \textit{La menzogna diventa verit\`a e passa alla storia}). We categorized pages according to their contents and their self description. The resulting dataset --downloaded over a timespan of four years (2010 to 2014)-- is composed of $73$ public Italian Facebook pages and it is the same used in~\cite{bessi2014science} and~\cite{bessi2014social}. To the best of our knowledge, the final dataset is the complete set of all scientific and conspiracy information sources active in the Italian Facebook scenario. Table~\ref{tab2} summarizes the details of our data collection.

\begin{table}[h]
\caption{ \textbf{Breakdown of the Facebook dataset.}}
	\begin{tabular}{|l|c|c|c|}
		\hline\bf {  }  & \bf {Total} & \bf {Science} & \bf {Conspiracy}  \\ 
		\hline 
		Pages & $ 73 $ & $ 34 $ & $ 39 $ \\ \hline
		Posts & $ 270,666 $ & $ 62,075 $ & $ 208,591 $ \\ \hline
		Likes & $ 9,164,781 $ & $ 2,505,399 $ & $ 6,659,382$  \\ \hline
		Comments & $1,017,509 $ & $ 180,918 $ & $ 836,591 $\\ \hline
		Shares &  $17,797,819$ & $1,471,088$ & $16,326,731$\\ \hline
		Likers & $ 1,196,404 $ & $ 332,357 $ & $ 864,047 $\\ \hline
		Commenters & $ 279,972 $ & $ 53,438 $ & $ 226,534 $\\ \hline
\end{tabular}\newline
\label{tab2}
\end{table}

%\begin{figure}[h]
%\includegraphics[width=0.6\textwidth]{fig/2.pdf}
%\caption{{\bf Figure Title first bold sentence Nulla mi mi, venenatis sed ipsum varius, volutpat euismod diam.}
%Figure Caption Proin rutrum vel massa non gravida. Quisque tempor sem et dignissim rutrum. A: Lorem ipsum dolor sit amet. B: Consectetur adipiscing elit.}
%\label{fig2}
%\end{figure}

\subsection*{Classification and annotator agreement metrics}

Our approach to sentiment classification of texts is based on supervised machine learning, 
where a sample of texts is first manually annotated with sentiment and then used to train and evaluate a classifier.
The classifier is then applied to the whole corpus.
The metrics to assess the agreement between annotators and the quality of the classifier
are based on contingency tables and confusion matrices, respectively. 

Annotators were asked to label each text with \textit{negative} $\prec$ \textit{neutral} $\prec$ \textit{positive} sentiment.
When two annotators are given the same text, they can either agree 
(both give the same label) or disagree (they give different labels). 
The annotators can disagree in two ways: one label is \textit{neutral} while the other is extreme (\textit{negative} or \textit{positive}), or both are extreme: one \textit{negative} and one \textit{positive} --- we call this severe disagreement. 
A convenient way to represent the overall (dis)agreement between the 
annotators is a contingency table,
where each text that is annotated twice appears in the table twice.
Table~\ref{generalCM} gives a generic $3\times3$ annotator agreement table,
while the actual data is in Table~\ref{AnnotatorAgreement}.
All agreements are on the diagonal of the table.
As the labels are ordered (\textit{negative} $\prec$ \textit{neutral} $\prec$ \textit{positive}), the further the cell from the diagonal, the more severe is the error.
From such a table one can calculate the annotator agreement (the sum of the main diagonal divided by the number of all the elements in the table) and 
the severe disagreement: the sum of top right and bottom left corners divided by the number of all the elements in the table.

To compare the predictions of a classifier to a golden standard (manually annotated data, in our case), a confusion matrix is used.
Table~\ref{generalCM} also represents a generic $3\times3$ confusion matrix for the (ordered) sentiment classification case. 
Each element $\langle x,y \rangle$ represents the number of examples
from the actual class $x$, predicted as class $y$.
All agreements/correct predictions are in the diagonal of the table.
In the ordinal classification case, the further the cell from the diagonal, the more severe is the error.

\begin{table}[h]
\caption{ \textbf{A generic $3\times3$ contingency table/confusion matrix.}}
\begin{tabular}{|l|c|c|c||c|}
\hline
\textbf{}         & \textit{Negative}   & \textit{Neutral}    & \textit{Positive}   & \textbf{Total}      \\ \hline
\textit{Negative} & $\langle -,- \rangle$ & $\langle -,0 \rangle$ & $\langle -,+ \rangle$ & $\langle -,* \rangle$ \\ \hline
\textit{Neutral}  & $\langle 0,- \rangle$ & $\langle 0,0 \rangle$ & $\langle 0,+ \rangle$ & $\langle 0,* \rangle$ \\ \hline
\textit{Positive} & $\langle +,- \rangle$ & $\langle +,0 \rangle$ & $\langle +,+ \rangle$ & $\langle +,* \rangle$ \\ \hline \hline
\textbf{Total}    & $\langle *,- \rangle$ & $\langle *,0 \rangle$ & $\langle *,+ \rangle$ & $N$ \\ \hline
\end{tabular}
\label{generalCM}
\end{table}

\acc\, is the fraction of correctly classified examples:
$$\mathit{Accuracy} = \frac{\langle-,-\rangle+\langle0,0\rangle+\langle+,+\rangle}{N}$$

%$$\mathit{Accuracy} = \frac{C_{\langle-,-\rangle} + C_{\langle0,0\rangle} + C_{\langle+,+\rangle}}{C_{\langle*,*\rangle}}$$

\textit{Accuracy within n} \cite{gaudette2009evaluation} allows for a wider range of predictions to be considered correct. 
We use \textit{Accuracy within 1} (\accone) where only misclassifications from \textit{negative} to \textit{positive} and vice-versa are considered incorrect:
$$\mathit{Accuracy}\!\pm\!1(-,+) = 1 - \frac{\langle+,-\rangle+\langle-,+\rangle}{N}$$

$\overline{F_1}(+,-)$ is the macro-averaged $F$-score of the positive and negative classes, a standard evaluation measure~\cite{kiritchenko2014sentiment} used also in 
the SemEval competition\footnote{SemEval: \url{http://alt.qcri.org/semeval2015/}} for sentiment classification tasks:
$$ \overline{F_1}(+,-) = \frac{\mathit{F_1}_+ + \mathit{F_1}_-}{2} $$

$F_1$ is the harmonic mean of Precision and Recall for each class
\cite{sokolova2009systematic}:
$$F_1 = 2 \cdot \frac{\mathit{Precision} \cdot \mathit{Recall}} {\mathit{Precision} + \mathit{Recall}} $$

Precision for class $x$ is the fraction of correctly
predicted examples out of all the predictions with class $x$:
$$Precision_x = \frac{\langle x,x\rangle}{\langle *,x\rangle} $$

Recall for class $x$ is the fraction of correctly predicted examples
out of all the examples with actual class $x$:
$$Recall_x = \frac{\langle x,x\rangle}{\langle x,*\rangle} $$

From the above tables and definitions, one can see that the annotator agreement is equivalent to \acc\, and that severe disagreement is equivalent to $1-$\accone.
$\overline{F_1}$ has no counterpart between the annotator agreement metrics,
but is a standard measure in evaluation of sentiment classifiers.

\subsection*{Data annotation}

Data annotation is a process in which some predefined labels are assigned to each data point. 
A subset of 19,642 comments from the Facebook dataset (Table~\ref{tab2}) was selected for manual sentiment annotation and later used to train a sentiment classifier. 
A user-friendly web and mobile devices annotation platform 
Goldfinch\footnote{provided by Sowa Labs: \url{http://www.sowalabs.com/}}
was used.

Trustworthy Italian native speakers, active on Facebook, were 
engaged for the annotations. 
The annotation task was to label each Facebook comment---isolated from its context---as  \textit{negative}, \textit{neutral}, or \textit{positive}.
The guideline given to the annotators was to estimate the emotional
attitude of the user when posting a comment to Facebook.
The exact question an annotator should answer was:
`Is the user happy (pleased, satisfied), or unhappy (angry, sad, frustrated), 
or neutral?'
%A dedicated Facebook group was formed to dispatch detailed annotation instructions, to provide a forum for discussion and to post ongoing annotation results that stimulated the annotators for contributing. No compensation, other then gratitude and personal satisfaction for having contributed to a good cause, were rewarded.
During the annotation process, which lasted for about two months, the annotator performance was monitored in terms of inter-annotator agreement and self-agreement,
based on 20\% of the comments which were intentionally duplicated.

The annotation process resulted in 23,675 annotated comments, 3,902 of them duplicated.
3,262 of them were polled to two different annotators and are used to
assess the inter-annotator agreement. 
The contingency table with the inter-annotator agreement 
is in Table~\ref{AnnotatorAgreement}. 
Note that, in the contingency table, each annotated example appears twice
(once for each of the two annotators), thus the matrix is symmetric. 
This is in contrast to a confusion matrix where one knows the ground
truth, and the matrix values are the numbers of examples in the actual
and predicted classes.

The three evaluation metrics described above were used to quantify the 
inter-annotator agreement. The results are in Table~\ref{tab0}.

%Accuracy($-,0,+ $)~=~72.0\%, 
%Accuracy\pm1($-,+$)~=~97.0\% 
%and $\overline{F1}(+,-)$~=~0.73.
%This means that the annotators gave the same labels to 72.0\% of the cases, while they severely disagree (one says \textit{positive} and the other says \textit{negative}) in 3\% of the cases.
%Accuracy is the percentage of examples on which the annotators agree (i.e., the sum of the diagonal of the confusion matrix divided by the number of all the examples in the confusion matrix). 
%Accurracy($-,0,+$) = 72.0\%, $F_{1}(-,+)$  = 3.0\%, ExtremeError($-,+$) = 0.73

\begin{table}[h]
\caption{ \textbf{A contingency table for the inter-annotator agreement, excluding self-agreement.}}
\begin{tabular}{|l|r|r|r||r|}
\hline
\textbf{}         & \textit{Negative} & \textit{Neutral} & \textit{Positive} & \textbf{Total} \\ \hline
\textit{Negative} & 2,482            & 545              & 90             & 3,117           \\ \hline
\textit{Neutral}  & 545              & 1,474             & 277               & 2,296          \\ \hline
\textit{Positive} & 90               & 277              & 744                & 1,111           \\ \hline \hline
\textbf{Total}    & 3,117             & 2,296            & 1,111             & 6,524           \\ \hline
\end{tabular}
\label{AnnotatorAgreement}
\end{table}

\subsection*{Classification}

Ordinal classification, also known as ordinal regression, is a form of multi-class classification where there is a natural ordering between the classes, 
but no meaningful numeric difference between them \cite{gaudette2009evaluation}.
%In this type of scenario, some errors are worse then others; in the case of sentiment analysis, a misclassification from \textit{positive} to \textit{negative} is worse compared to a misclassification from \textit{positive} to \textit{neutral}. Besides the usual quality metics for multi-class classification, specific metrics like \accone\ \cite{gaudette2009evaluation} and $\overline{F1}(+,-)$ \cite{kiritchenko2014sentiment} have been defined to properly assess the quality of an ordinal classifier.
We treat sentiment classification as an ordinal regression task
with three ordered classes.
We apply the wrapper approach, described in \cite{frank2001simple}, with two linear-kernel Support Vector Machine (SVM) \cite{vapnik95} classifiers.
SVM is a state-of-the-art supervised learning algorithm,
well suited for large scale text categorization tasks,
and robust on large feature spaces. 
The two SVM classifiers were trained to distinguish the extreme classes 
(\textit{negative} and \textit{positive}) from the rest
(\textit{neutral} plus \textit{positive}, and \textit{neutral} plus \textit{negative},
respective).
During prediction, if both classifiers agree, they yield the common class,
otherwise, if they disagree, the assigned class is \textit{neutral}.

The sentiment classifier was trained and tuned on the train set of 15,714 annotated comments. The comments were processed into the standard Bag-of-Words (BoW) 
representation, with the following settings: lemmatized BoW include unigrams and bigrams, minimum n-gram frequency is five, TF-IDF weighting, no stop-word removal, and normalized vectors. Additional features and settings were chosen, based on
the results of 10-fold stratified cross-validation on the train set:
normalization of diacritical characters, url replacement, length of text,
presence of upper cased words, negation (language specific), swearing (language specific), positive words from a predefined dictionary (language specific), unusual punctuation (several exclamation or question marks, ...), unusually repeated characters, happy or sad emoticons in the text, and their presence at the end of the sentence.

The trained sentiment classifier was then evaluated on a 
disjoint test set of 3,928 comments.
The confusion matrix between the annotators (actual classes) and the classifier 
(predicted classes) is in Table~\ref{CMTestSet}. 
The evaluation results on the test set are in Table~\ref{tab0}.
The sentiment class distribution, after applying the classifier to the whole
set of Facebook comments, is in Figure~\ref{fig0}.

\begin{table}[h]
\caption{ \textbf{Confusion matrix of the sentiment classifier on the test set.}}
\begin{tabular}{|l|r|r|r||r|}
\hline
\backslashbox{Actual}{Predicted} & \textit{Negative} & \textit{Neutral} & \textit{Positive} & \textbf{Total} \\ \hline
\textit{Negative} & 1,208                & 545              & 90             & 1,747         \\ \hline
\textit{Neutral}  & 509               & 987              & 103               & 1,599         \\ \hline
\textit{Positive} & 86               & 183              & 319                & 588            \\ \hline \hline
\textbf{Total}    & 454               & 1,671            & 1,803             & 3,928           \\ \hline
\end{tabular}
\label{CMTestSet}
\end{table}

\subsection*{Statistical tools}

To characterize random variables, a main tool is the probability distribution
function (PDF), which gives the probability that a random variable
$X$ assumes a value in the interval $[a,b]$, i.e. $P(a \leq X \leq b) = \int_{a}^{b} f(x) dx$.
%The cumulative distribution function (CDF) is another important tool
%giving the probability that a random variable $X$ is less than or
%equal to a given value $x$, i.e. $F(x) = P(X \leq x) = \int_{-\infty}^{x}f(y)dy$.
%In social sciences, an often occuring probability distribution function
%is the Pareto's law $f(x) \sim x^{-\gamma}$, that is characterized
%by power law tails, i.e. by the occurrence of rare but relevant events.
%In fact, while $f(x) \to 0$ for $x \to \infty$ (i.e. high values
%of a random variable $X$ are rare), the total probability of rare
%events is given by $C(x) = P(X > x) = \int_{x}^{\infty}f(y)dy$, where
%$x$ is a sufficiently large value. Notice that $C(x)$ is the Complement
%to the CDF (CCDF), where complement indicates that $C(x) = 1 - F(x)$. Hence, in order to better visualize the behavior
%of empirical heavy--tailed distributions, we recur to log--log plots
%of the CCDF.

\begin{figure}[H]
 \centering
       \includegraphics[width=.55\textwidth]{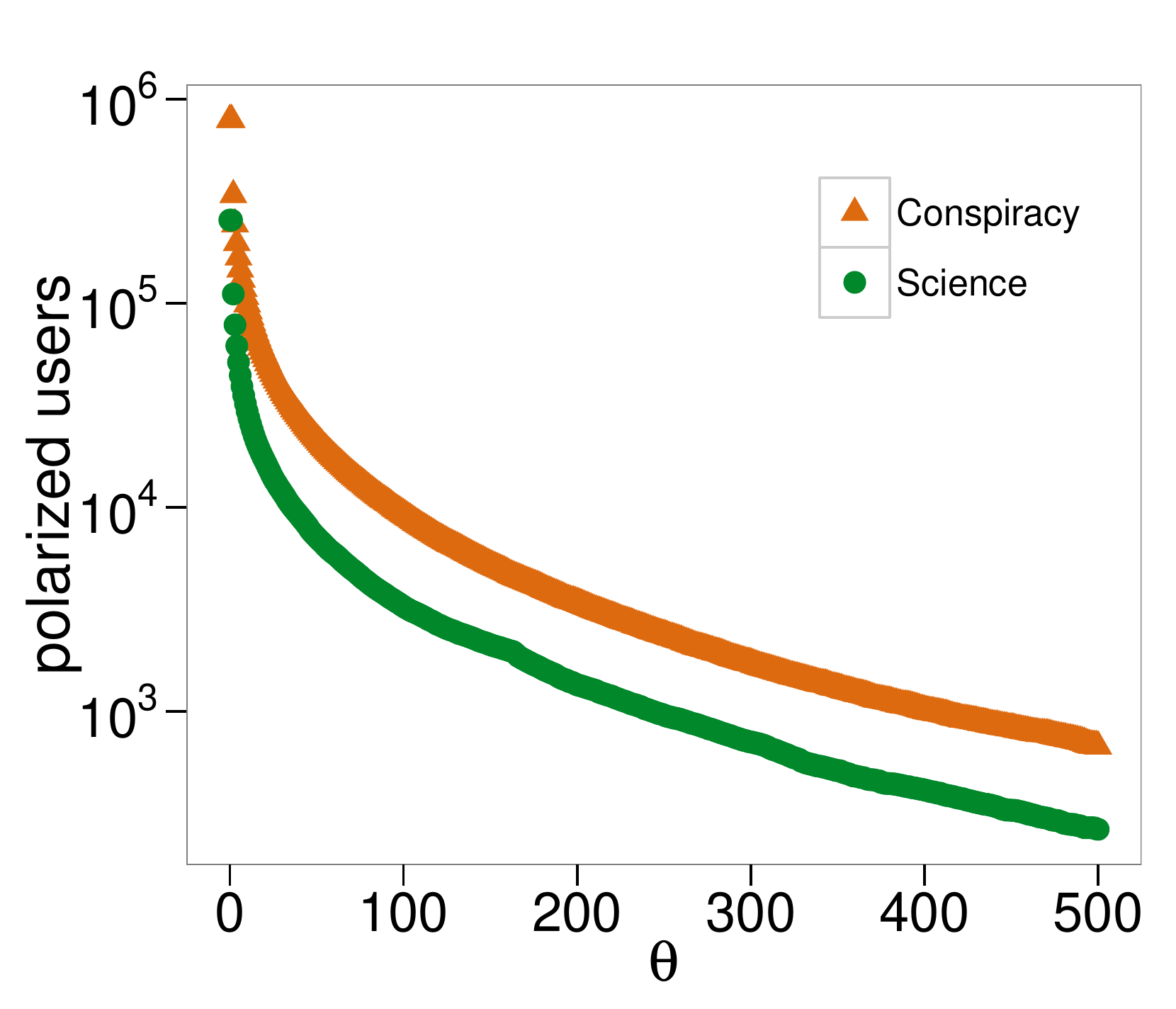}
\caption{\textbf{Polarized users and activity.} The number of polarized users as a function of the thresholding value $\theta$ on the two categories.}
\label{fig:tresholding}
\end{figure}

\begin{figure}[H]
 \centering
       \includegraphics[width=.70\textwidth]{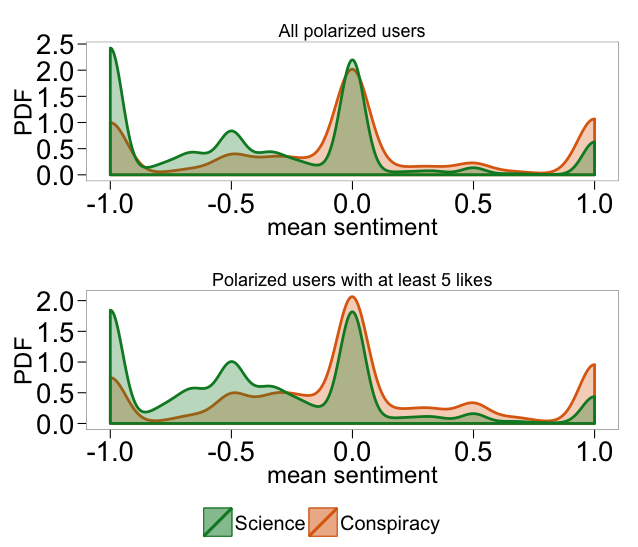}
\caption{\textbf{Sentiment of Polarized Users.} Probability Density Function (PDF) of the mean sentiment of all polarized users (top) and polarized users with at least five likes, where $-1$ corresponds to negative sentiment, $0$ to neutral and $1$ to positive.}
\label{fig:polarized_sentiment}
\end{figure}

\begin{figure}[H]
 \centering
       \includegraphics[width=.55\textwidth]{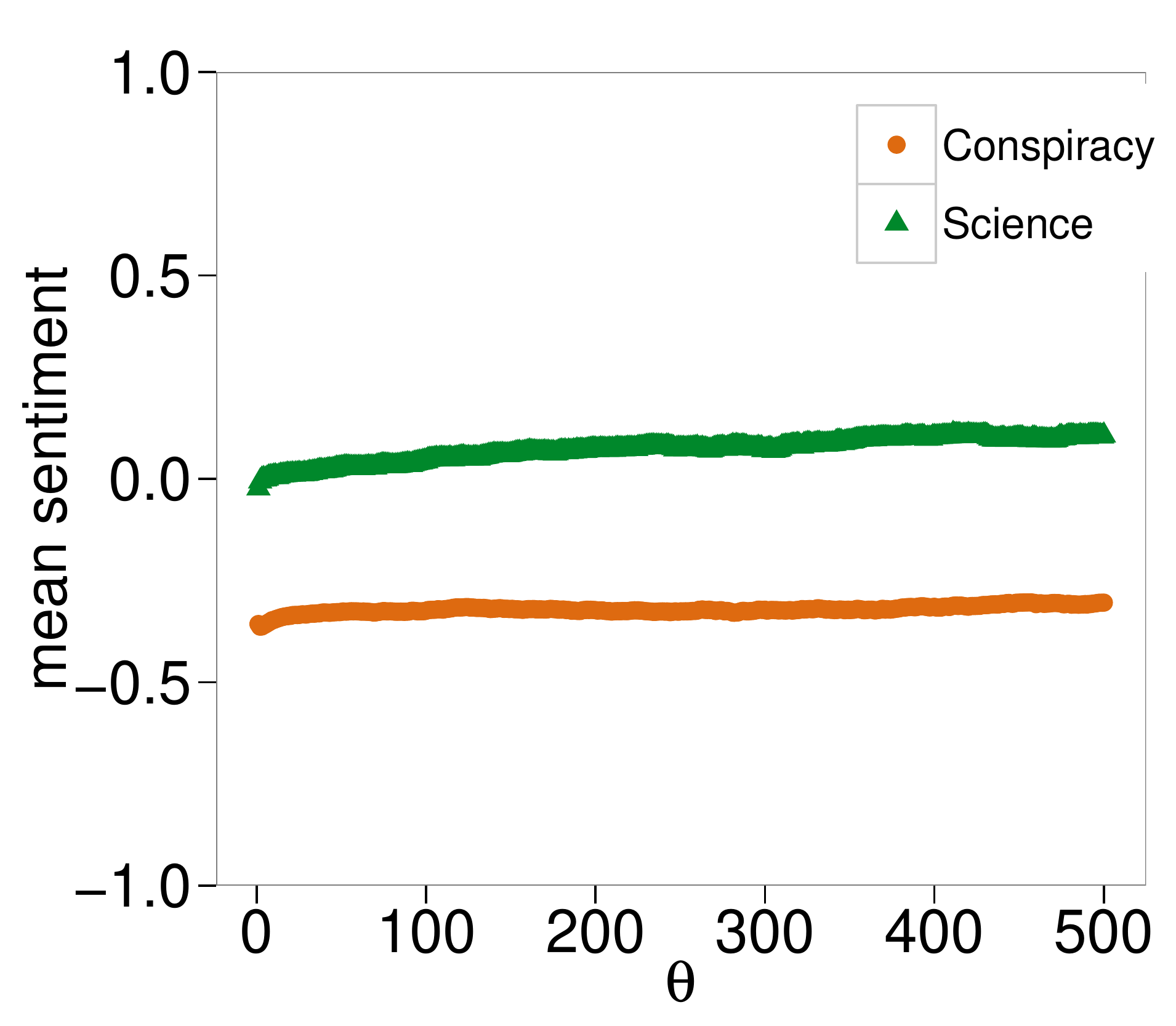}
\caption{\textbf{Sentiment and Engagement.} Average sentiment of polarized users as a function of the threshold $\theta$, i.e., the engagement degree, intended as the number of likes a polarized user put in her own category.}
\label{fig:polarized_sentiment2}
\end{figure}

\paragraph{Labeling algorithm.}
The labeling algorithm may be described as a thresholding strategy on the total number of users likes. Considering the total number of likes of a user $L_u$ on both posts $P$ in categories $S$ and $C$. Let  $l_s$ and $l_c$ define the number of likes of a user $u$ on $P_s$ or $P_c$, respectively denoting posts from scientific or conspiracy pages.  Then, the total like activity of a user on one category is given by $\frac{l_s}{L_u}$. Fixing a threshold $\theta$ we can discriminate users with enough activity on one category. More precisely, the condition for a user to be labeled as a polarized user in one category can be described as $\frac{l_s}{L_u}$ $\vee$ $\frac{l_c}{L_u}$ $> \theta$. 
In Fig.~\ref{fig:tresholding} we show the number of polarized users as a function of $\theta$. Both curves decrease with a comparable rate. Fig.~\ref{fig:polarized_sentiment} shows the Probability Density Function (PDF) of the mean sentiment of all polarized users \textit{(top)} and polarized users with at least five likes \textit{(bottom)}. Note that both densities are qualitatively similar. In Fig.~\ref{fig:polarized_sentiment2} we show the mean sentiment of polarized users as a function of the threshold $\theta$. 

\subsection*{List of pages}
\label{sec:page_list}

In this section, we provide the full list of Facebook pages of our dataset. Table~\ref{tab3} lists scientific pages, while Table~\ref{tab4} lists conspiracy pages.

\begin{table}[H]
\begin{footnotesize}
\centering
\begin{tabular}{|r|l|l|}
  \hline
 & \textbf{Page Name} & \textbf{Link} \\ 
  \hline
 1 & Scientificast.it & www.facebook.com/129133110517884 \\ 
  2 & CICAP & www.facebook.com/32775139194 \\ 
  3 & OggiScienza & www.facebook.com/106965734432 \\ 
  4 & Query & www.facebook.com/128523133833337 \\ 
  5 & Gravit\`a Zero & www.facebook.com/138484279514358 \\ 
  6 & COELUM Astronomia & www.facebook.com/81631306737 \\ 
  7 & MedBunker & www.facebook.com/246240278737917 \\ 
  8 & In Difesa della Sperimentazione Animale & www.facebook.com/365212740272738 \\ 
  9 & Italia Unita per la Scienza & www.facebook.com/492924810790346 \\ 
  10 & Scienza Live & www.facebook.com/227175397415634 \\ 
  11 & La scienza come non l'avete mai vista & www.facebook.com/230542647135219 \\ 
  12 & LIBERASCIENZA & www.facebook.com/301266998787 \\ 
  13 & Scienze Naturali & www.facebook.com/134760945225 \\ 
  14 & Perch\'e vaccino & www.facebook.com/338627506257240 \\ 
  15 & Le Scienze & www.facebook.com/146489812096483 \\ 
  16 & Vera scienza & www.facebook.com/389493082245 \\ 
  17 & Scienza in rete & www.facebook.com/84645527341 \\ 
  18 & Galileo, giornale di scienza e problemi globali & www.facebook.com/94897729756 \\ 
  19 & Scie Chimiche: Informazione Corretta & www.facebook.com/351626174626 \\ 
  20 & Complottismo? No grazie & www.facebook.com/399888818975 \\ 
  21 & INFN - Istituto Nazionale di Fisica Nucleare & www.facebook.com/45086217578 \\ 
  22 & Signoraggio: informazione corretta & www.facebook.com/279217954594 \\ 
  23 & JFK informazione corretta & www.facebook.com/113204388784459 \\ 
  24 & Scetticamente & www.facebook.com/146529622080908 \\ 
  25 & Vivisezione e Sperimentazione Animale, verit\`a e menzogne & www.facebook.com/548684548518541 \\ 
  26 & Medici Senza Frontiere & www.facebook.com/65737832194 \\ 
  27 & Task Force Pandora & www.facebook.com/273189619499850 \\ 
  28 & VaccinarSI & www.facebook.com/148150648573922 \\ 
  29 & Lega Nerd & www.facebook.com/165086498710 \\ 
  30 & Super Quark & www.facebook.com/47601641660 \\ 
  31 & Curiosit\`a Scientifiche & www.facebook.com/595492993822831 \\ 
  32 & Minerva - Associazione di Divulgazione Scientifica & www.facebook.com/161460900714958 \\ 
  33 & Pro-Test Italia & www.facebook.com/221292424664911 \\ 
  34 & Uniti per la Ricerca & www.facebook.com/132734716745038 \\ 
   \hline
\end{tabular}
\caption{\textbf{Scientific news sources}: List of Facebook pages diffusing main stream scientific news and their url.}
\label{tab3}
\end{footnotesize}
\end{table}

\begin{table}[H]
\begin{footnotesize}
\centering
\begin{tabular}{|r|l|l|}
  \hline
 & \textbf{Page Name} & \textbf{Link} \\ 
  \hline
1 & Scienza di Confine & www.facebook.com/188189217954979 \\ 
  2 & CSSC - Cieli Senza Scie Chimiche & www.facebook.com/253520844711659 \\ 
  3 & STOP ALLE SCIE CHIMICHE & www.facebook.com/199277020680 \\ 
  4 & Vaccini Basta & www.facebook.com/233426770069342 \\ 
  5 & Tanker Enemy & www.facebook.com/444154468988487 \\ 
  6 & SCIE CHIMICHE & www.facebook.com/68091825232 \\ 
  7 & MES Dittatore Europeo & www.facebook.com/194120424046954 \\ 
  8 & Lo sai & www.facebook.com/126393880733870 \\ 
  9 & AmbienteBio & www.facebook.com/109383485816534 \\ 
  10 & Eco(R)esistenza & www.facebook.com/203737476337348 \\ 
  11 & curarsialnaturale & www.facebook.com/159590407439801 \\ 
  12 & La Resistenza & www.facebook.com/256612957830788 \\ 
  13 & Radical Bio & www.facebook.com/124489267724876 \\ 
  14 & Fuori da Matrix & www.facebook.com/123944574364433 \\ 
  15 & Graviola Italia & www.facebook.com/130541730433071 \\ 
  16 & Signoraggio.it & www.facebook.com/278440415537619 \\ 
  17 & Informare Per Resistere & www.facebook.com/101748583911 \\ 
  18 & Sul Nuovo Ordine Mondiale & www.facebook.com/340262489362734 \\ 
  19 & Avvistamenti e Contatti & www.facebook.com/352513104826417 \\ 
  20 & Umani in Divenire & www.facebook.com/195235103879949 \\ 
  21 & Nikola Tesla - il SEGRETO & www.facebook.com/108255081924 \\ 
  22 & Teletrasporto & www.facebook.com/100774912863 \\ 
  23 & PNL e Ipnosi & www.facebook.com/150500394993159 \\ 
  24 & HAARP - controllo climatico & www.facebook.com/117166361628599 \\ 
  25 & Sezione Aurea, Studio di Energia Vibrazionale & www.facebook.com/113640815379825 \\ 
  26 & PER UNA NUOVA MEDICINA & www.facebook.com/113933508706361 \\ 
  27 & PSICOALIMENTARSI E CURARSI NATURALMENTE & www.facebook.com/119866258041409 \\ 
  28 & La nostra ignoranza  la LORO forza. & www.facebook.com/520400687983468 \\ 
  29 & HIV non causa AIDS & www.facebook.com/121365461259470 \\ 
  30 & Sapere  un Dovere & www.facebook.com/444729718909881 \\ 
  31 & V per Verit\`a & www.facebook.com/223425924337104 \\ 
  32 & Genitori veg & www.facebook.com/211328765641743 \\ 
  33 & Operatori di luce & www.facebook.com/195636673927835 \\ 
  34 & Coscienza Nuova & www.facebook.com/292747470828855 \\ 
  35 & Aprite Gli Occhi & www.facebook.com/145389958854351 \\ 
  36 & Neovitruvian & www.facebook.com/128660840526907 \\ 
  37 & CoscienzaSveglia & www.facebook.com/158362357555710 \\ 
  38 & Medicinenon & www.facebook.com/248246118546060 \\ 
  39 & TERRA REAL TIME & www.facebook.com/208776375809817 \\ 
   \hline
\end{tabular}
\caption{\textbf{Conspiracy news sources}: List of Facebook pages diffusing conspiracy news and their url.}
\label{tab4}
\end{footnotesize}
\end{table}

%\section*{References}
% Either type in your references using
% \begin{thebibliography}{}
% \bibitem{}
% Text
% \end{thebibliography}
%
% OR
%
% Compile your BiBTeX database using our plos2015.bst
% style file and paste the contents of your .bbl file
% here.
% 

\bibliography{biblio_sent}

\end{document}